\input epsf

\documentstyle[eqsecnum,aps,prd]{revtex} 
\preprint{DAMTP R-97/05, UPENN 97/0731T} 
\tighten
\begin{document}
\draft
\def\sqr#1#2{{\vcenter{\hrule height.3pt
      \hbox{\vrule width.3pt height#2pt  \kern#1pt
         \vrule width.3pt}  \hrule height.3pt}}}
\def\square{\mathchoice{\sqr67\,}{\sqr67\,}\sqr{3}{3.5}\sqr{3}{3.5}}
\def\today{\ifcase\month\or
  January\or February\or March\or April\or May\or June\or
  July\or August\or September\or October\or November\or December\fi
  \space\number\day, \number\year}


\title{Cosmic Strings in an Open Universe: \\Quantitative Evolution
and Observational Consequences}

\author{P. P. Avelino${}^{1}$\thanks{ 
Electronic address: pedro\,@\,pulsar.astro.up.pt},
R. R. Caldwell${}^{2}$\thanks{ 
Electronic address: caldwell\,@\,dept.physics.upenn.edu}, 
and C. J. A. P. Martins${}^{3}$\thanks{Also at C. A. U. P.,
Rua do Campo Alegre 823, 4150 Porto, Portugal. 
Electronic address: C.J.A.P.Martins\,@\,damtp.cam.ac.uk}}

\address{${}^1$ Centro de Astrof\'{\i}sica, Universidade do Porto\\
Rua do Campo Alegre 823\\
PT-4150 Porto, Portugal}

\address{${}^2$ Department of Physics and Astronomy\\
University of Pennsylvania\\
Philadelphia, PA 19104}

\address{${}^3$ Department of Applied Mathematics and Theoretical Physics\\
University of Cambridge\\
Silver Street, Cambridge CB3 9EW, U.K.}

\maketitle
\begin{abstract}

The cosmic string scenario in an open universe is developed -- including the
equations of motion,  a model of network evolution, the large  angular scale
cosmic microwave background (CMB) anisotropy, and the power spectrum of density
fluctuations produced by cosmic strings with dark matter. We first derive the
equations of motion for cosmic string in an open Friedmann-Robertson-Walker
(FRW)  space-time.  With these equations and the  cosmic string stress-energy
conservation law, we construct a quantitative model of the evolution of the
gross features of a cosmic string network in a dust-dominated, $\Omega <1$ FRW
space-time. Second, we apply this model of network evolution to the results of a
numerical simulation of cosmic strings in a dust-dominated, $\Omega =1$ FRW
space-time, in order to estimate the rms temperature anisotropy induced by
cosmic strings in the CMB. By comparing  to the COBE-DMR observations, we obtain
the normalization for the cosmic string mass per unit length $\mu$ as a function
of $\Omega$.  Third, we consider the effects of the network evolution and
normalization in an open universe on the large scale structure formation
scenarios with either cold or hot dark matter (CDM, HDM). The string+HDM
scenario for $\Omega < 1$ appears to produce too little power on scales $k
\gtrsim 1\, \Omega h^2/$Mpc. In a low density universe the string+CDM scenario
is a better model for structure formation.  We find that for cosmological
parameters $\Gamma = \Omega h \sim 0.1 - 0.2$ in an open universe the string+CDM
power spectrum fits the shape of the linear power spectrum inferred from various
galaxy surveys. For $\Omega \sim 0.2 - 0.4$, the model requires a bias $b
\gtrsim 2$ in the variance of the mass fluctuation on scales $8 \, h^{-1}$Mpc. 
In the presence of a cosmological constant, the spatially-flat string+CDM power
spectrum requires a slightly lower bias than for an open universe of the same
matter density.

\end{abstract} 

\section{Introduction}
\label{secintro}

Cosmic strings are topological defects which may have formed in the very early
universe and may be responsible for the formation of large scale structure
observed in the universe today
\cite{VilenkinReport,VilenkinShellard,HKReport}.  In order to test the
hypothesis that the inhomogeneities in our universe were induced by cosmic
strings one must compare observations of our universe with the predictions of
the cosmic string model. To date, most work on the cosmic string scenario has
been  carried out with a background cosmological model which is a spatially
flat, $\Omega  = 1$ FRW space-time. (See \cite{Spergel,PenSpergel,Ferreira,LuoSchramm} for
other work on open scenarios with defects.) 

Observational evidence indicates that to within $95 \%$ confidence, the
present-day cosmological density parameter lies in the range $0.2 < \Omega  <
2$, and is most likely less than or equal to unity \cite{OpenEvidence}. This
reason alone is enough motivation  to investigate open cosmologies.  We are
further compelled when we recognize that the string+CDM linear power spectrum,
while producing too much small scale power in the case $\Omega = 1$
\cite{AlbrechtStebbins}, appears to fit the the shape of the power spectrum 
estimated from the various three-dimensional  galaxy redshift surveys for
$\Omega < 1$ \cite{PenSpergel,Ferreira}. Hence, we aim to develop the tools
necessary to study cosmic strings in an open universe.

The outline of this paper is as follows. In section \ref{secopen} we construct
a background cosmology composed of a dust-dominated, $\Omega  < 1$ FRW
space-time. We derive the cosmic string equations of motion and the energy
conservation equation in an open universe, and discuss the effects of the
spatial curvature and rapid, curvature-dominated expansion on the density of
strings through a simple solution of these equations. 

We next construct an analytic model of the long string evolution in an open
universe. While not as sophisticated as the model developed by
Austin-Copeland-Kibble \cite{ACK},  we improve on earlier work
\cite{SimpleModels} by following the procedure of Martins and Shellard (MS)
\cite{MartinsShellard}, which treats the mean string velocity as a dynamical
variable. We should point out that the MS model provides an accurate
description of the behaviour of a cosmic string network seen in numerical
simulations  of radiation- through matter-dominated expansion  in the case
$\Omega=1$. Since no such simulations exist in the case of an open universe,
the model presented in this paper is an extrapolation. Nevertheless, the onset
of curvature domination in the present case  is  qualitatively similar to the
radiation-matter transition;  in both cases, the dominant dynamical effect
during the transition is due to the shift in the time dependence of the scale
factor. Hence, we expect that the model developed in this section, which
includes the effects of curvature-driven expansion and spatial curvature on the
the string equations of motion, will be sufficient to provide a good
description of cosmic string evolution in an open universe -- although future
numerical work will undoubtedly be required to test this model.

In section \ref{secevol}, by numerically solving the evolution equations, we
find that with the onset of curvature domination, the mean string velocity and
energy density decay rapidly. We also note that there does not appear to be a
scaling solution for the gross features of the string network, as occurs in a
spatially-flat, $\Omega =1$ cosmology. Similar work has also been carried  out
recently by Martins \cite{Martins}. In section \ref{seccmb} we construct a  
semi-analytic model of the CMB anisotropy induced by strings in an open
cosmology. We obtain the normalization of the string mass per unit length
$\mu$, as a function of $\Omega$, by comparing with the COBE-DMR observations.
Next, we consider the effect of the new normalization  on the large scale
structure power spectrum when $\Omega  < 1$ by adapting Albrecht and Stebbins'
\cite{AlbrechtStebbins} semi-analytic model for the string+CDM and HDM
scenarios. While the power spectrum does not completely specify the
non-gaussian fluctuation patterns generated by cosmic string wakes, it serves
as a useful gauge of the viability of the scenario. We find that in an open
universe, the string+HDM  spectrum suffers from a lack of power on small
scales, compared to  the linear power spectrum estimated from  various galaxy
redshift surveys. However, the string+CDM spectrum in an open universe with
bias $b \gtrsim 2$ appears to fit the  observed power spectrum. Finally, in
section \ref{seccosmoconst}, we consider the case of a spatially flat, low
matter density universe with a cosmological constant. Applying the same tools
that we have developed for the study of an open universe, we obtain the CMB
normalization of the mass per unit length. We find that the string+CDM power
spectrum requires a slightly lower bias than for an open universe with the same
matter density. We conclude in section \ref{secend}.

Throughout this paper we have set the speed of light to unity, $c = 1$, and
adopted the convention  $H_o \equiv 100\, h \,{\rm km}\, {\rm sec}^{-1}\,{\rm
Mpc}^{-1}$.

\section{Open Universe}
\label{secopen}

To begin our investigation of cosmic strings in an open FRW space-time, we must
construct a background cosmology. We find it convenient to use the metric 
\begin{equation}
ds^2 = a^2(\tau)\Big[ d\tau^2 - (1 + K r^2)^{-2} 
\Big( dx^2 + dy^2 + dz^2 \Big) \Big]
\label{ourmetric}
\end{equation}
where $r^2 = (x^2 + y^2 + z^2)/R_c^2$, $K= -1$ for an open FRW space-time, and
the coordinates lie in the range $\{x,y,z\} \in (-R_c,R_c)$. Note that for
$K=-1,0,+1$ the spatial sections are ${\rm H}^3,\,{\rm R}^3,\,{\rm S}^3$
respectively. The radius of spatial curvature is given by
\begin{equation}
R = a(\tau) R_c = H^{-1} |1-\Omega|^{-1/2} . 
\label{curvscale}
\end{equation}
Thus, a universe with a density close to critical has a very large
radius of curvature and is very flat, whereas in a low density universe
the curvature radius is comparable to the Hubble radius.  The
cosmological time $t$ is related to the conformal time by $t = \int
d\tau a(\tau)$.

In this space-time, the induced metric on the cosmic string world sheet is
\begin{eqnarray}
\gamma_{ab} = g_{\mu \nu} {x^\mu}_{,a} {x^\nu}_{,b}
\rightarrow  
\gamma_{\tau\tau} &=& a^2(\tau)\Big[1 - {\dot {\bf x}^2
\over (1 + K r^2)^2}\Big] \cr
\gamma_{\sigma\sigma} &=& -a^2(\tau){{{\bf x}'}^2 \over (1 + K r^2)^2} .
\end{eqnarray}
The indices $a,b$ denote coordinates $\tau,\sigma$ on the string worldsheet,
with  $\dot{} = \partial_\tau$ and ${}' = \partial_\sigma$, where $\sigma$ is a
parameter along the string.  We have chosen the gauge such that the time
parameter along the string coincides with the conformal time $\tau$, and $\dot
{\bf x} \cdot {\bf x}' = 0$.

Referring to \cite{VilenkinShellard} (equation 6.1.15) for the string
stress-energy tensor, we find that the energy density in string is given by
\begin{eqnarray}
\rho &=& {\mu \over a^2} \int d\sigma \,
\sqrt{-\gamma} \gamma^{ab} {x^\tau}_{,a} {x^\tau}_{,b} \,
\delta^3({\bf x} - {\bf x}(\tau,\sigma)) \cr
&=& {\mu \over a^2} \int d\sigma \,
{{\bf x}' \over \sqrt{(1 + K r^2)^2 - \dot {\bf x}^2}} 
\,\delta^3({\bf x} - {\bf x}(\tau,\sigma)) 
\cr
&=& {\mu \over a^2} \int d\sigma \, \tilde \epsilon \,
\delta^3({\bf x} - {\bf x}(\tau,\sigma))
\qquad \Rightarrow \tilde\epsilon \equiv
{{\bf x}' \over \sqrt{(1 + K r^2)^2 - \dot {\bf x}^2}}.
\end{eqnarray}
Here, $\tilde\epsilon$ is the comoving coordinate length of string per unit
$\sigma$. The integral above tells us that the total energy in string is not
simply the total length of string, weighted by a relativistic $\gamma$ factor,
times the mass per unit length $\mu$. The spatial curvature also has an effect,
as there is a contribution to the energy by the $Kr^2-$term for strings with large
spatial extent. This coupling of curvature to the string energy will have an
important effect on the evolution of very large strings.

The equations of motion for cosmic string in the space-time
(\ref{ourmetric}) are given by
(generalizing equation 6.1.12 of \cite{VilenkinShellard})
\begin{eqnarray}
&& {x^\mu_{,a}}^{;a} + \Gamma^\mu_{\nu\rho} 
\gamma^{ab} x^\nu_{,a}x^\rho_{,b}
 = 0 \cr\cr \mu=\tau \quad \rightarrow \quad &&
\dot{\tilde\epsilon} = -2 {\dot a \over a} \tilde \epsilon  
{\dot {\bf x}^2  \over (1 + K r^2)^2} 
\label{conservationeqn}\\
\mu = i \quad \rightarrow \quad   &&
\ddot x^i + 2 {\dot a \over a} \dot x^i
\Big(1 - {\dot {\bf x}^2 \over (1 + K r^2)^2} 
\Big) - {1 \over \tilde \epsilon} \Big( {{x}'^i \over \tilde \epsilon} \Big)'
\cr\cr && \qquad
= {2 K a^2 H^2 |1 - \Omega| \over 1 + K r^2}
\Big[
x^i ( \dot {\bf x}^2 - \tilde \epsilon^{-2}{\bf x}'^2 )
- 2 \dot x^i  ({\bf x} \cdot \dot {\bf x})
+ 2 \tilde \epsilon^{-2} {x}'^i  ({\bf x} \cdot {\bf x}')
\Big]. 
\label{openeqnofmotion}
\end{eqnarray}
The above equations represent the first main result of this paper;
by setting $K=-1$, (\ref{conservationeqn}) and (\ref{openeqnofmotion})
give the equations of motion for cosmic strings in an open FRW
space-time. Similarily, $K=+1,\,0$ give
the equations of motion in closed and flat FRW space-times.
Hereafter we will only consider the case of an open universe.

We can get a better idea of the effects of the rapid expansion and
space-time curvature by studying a simple solution of these microscopic
equations.  For our purposes it is sufficient to consider dust-dominated expansion only, for which the scale factor is given by
\begin{equation}
a(\tau) = {\Omega \sinh^2(\tau/2) \over H_o (1 - \Omega)^{3/2}}.
\label{dustfactor}
\end{equation}
Here $H_o,\,\Omega$ are the present-day Hubble constant and
cosmological density parameter.  We have solved (\ref{openeqnofmotion})
for the case of a circular loop with an initial physical radius
$R_{loop}(t_{eq})=10\,t_{eq}$ and velocity $v(t_{eq})=0$ at the time of
radiation-matter equality. The evolution of the loop radius and
velocity, for the cases $\Omega =1,\, 0.2$ are shown in Fig.
\ref{figure0}. At early times the evolution is indistinguishable, as
the loop is conformally stretched by the expansion and picks up speed.
As the loop falls inside the Hubble horizon, it begins to oscillate.
We see that the frequency of oscillation, while constant relative to
$H$, is higher relative to $t^{-1}$ in an open universe owing to the
faster expansion rate. This is crucial because oscillating loops in an
expanding universe lose a constant fraction of their energy in each
period \cite{MartinsShellard,Garriga}. Hence, the average velocity at
late times is smaller than the spatially-flat value $v^2_{flat}=1/2$.
Incidentally, the velocity damping guarantees that the denominator of
$\tilde\epsilon$, $(1 - r^2)^2 - \dot x^2$, is always positive; for
large $r$, $\dot x$ must decrease.  In short, the curvature-driven
expansion serves to damp the string motion.

We will move on to concentrate on (\ref{conservationeqn}), the
conservation equation, for the present. Making the definition
\begin{equation}
v^2 \rho = 
{\mu \over a^2} \int d\sigma \,
\tilde \epsilon  \dot {\bf x}^2 \,
\delta^3({\bf x} - {\bf x}(\tau,\sigma)),
\label{veldef}
\end{equation}
for the mean squared string velocity, we obtain the energy density
conservation equation
\begin{equation}
\dot \rho + 2 {\dot a \over a}\rho =
-2 {\dot a \over a} {\mu \over a^2} \int d\sigma \, \tilde \epsilon
{\dot {\bf x}^2 \over (1 - r^2)^2}\,
\delta^3({\bf x} - {\bf x}(\tau,\sigma)). 
\label{rhoevolution}
\end{equation}
For $r \to 1$ ($0< r^2 < 1$) with fixed $\dot {\bf x}^2$  the term on
the RHS of (\ref{rhoevolution}) becomes large and negative. Hence, this
confirms that the effect of the spatial  curvature will be to enhance
the dilution of the energy density in long strings due to the
expansion.

Let us examine (\ref{rhoevolution}) more closely. The  integrand on the
RHS is just the coordinate energy $\tilde \epsilon$  times the velocity
squared, weighted by a factor   of the ratio $r$ of the string length
to the curvature scale.  By averaging over strings of length  scale
$\lambda$, we may rewrite (\ref{rhoevolution}) as
\begin{equation}
\dot \rho_\lambda + 2 {\dot a \over a} \rho_\lambda  = 
-2 {\dot a \over a} \rho_\lambda  \langle v^2 \rangle 
\Big[1 - (\lambda/R)^2\Big]^{-2}.
\label{fevolution}
\end{equation}
The ratio $\lambda/R$ determines the importance of the curvature
contribution.  For strings much smaller than the curvature scale,
$\lambda \ll R$, we obtain the usual flat space energy density
conservation equation. For very large strings, $\lambda \to R$. Hence,
string with support on very large scales samples more of, or is more
tightly coupled to the curvature. Crudely, the effect is that the long
string energy density is dissipated more rapidly as the space-time
expands.

\section{Quantitative String Evolution} \label{secevol}

We now build on the work of Martins and Shellard \cite{MartinsShellard} to
construct an analytic model of long string network evolution in an open
universe. As carried out by MS, we treat the average string velocity as well as
the characteristic string length scale as dynamical variables. For long
strings, the characteristic length scale is related to the network density of
long strings by $\rho_L = \mu L^{-2}$. Hence, we obtain from (\ref{fevolution})
\begin{equation}
{d L \over d t} = LH \Big(1 + v^2 \Big[ 1 - (1-\Omega)(LH)^2 \Big]^{-2} \Big)
+ {1 \over 2} \tilde c v.
\label{Levolutioneqn}
\end{equation}
The phenomenological loop chopping efficiency parameter $\tilde c$ models the
transfer of energy from the long strings to loops. Next, an evolution equation
for the velocity may be obtained by differentiating (\ref{veldef}) and using
(\ref{openeqnofmotion}):
\begin{equation}
{d v \over d t} = \Big(\Big[ 1 - (1-\Omega)(LH)^2 \Big]^2- v^2\Big) {\kappa \over L} - 2 H v 
\Big(1 - v^2\Big[ 1 - (1-\Omega)(LH)^2 \Big]^{-2} \Big) \,.
\label{vevolutioneqn}
\end{equation} 
As in MS ({\it c.f.} equations 2.40 and 2.41 of \cite{MartinsShellard}) the
parameter $\kappa$  (MS use $k$) has been introduced to describe  the presence
of small scale structure on the long strings.  Equations
(\ref{Levolutioneqn}-\ref{vevolutioneqn}) are the second main result of this
paper. Again, in the limit $\Omega \to 1$, the flat space evolution equations
(equations 2.20 and 2.38 of \cite{MartinsShellard}) are obtained.  (Note that
the above equations are equally valid in a closed, $\Omega > 1$ FRW
space-time.) We have omitted the friction damping terms due to the interaction
of the cosmic strings with the hot cosmological fluid, which are important only
near the time the strings were formed. Similar equations are given in Ref.
\cite{Martins}, although it was assumed that the contribution of the curvature
terms is negligible.

In a spatially flat, $\Omega  = 1$ FRW space-time, scaling solutions may be
found for which $\dot v = 0$ and $L/t=$constant. In a dust-dominated era, the
Allen-Shellard numerical simulation suggests the values $\kappa = 0.43$ and
$\tilde c = 0.15$ (note that MS used $\kappa = 0.49$ and $\tilde c=0.17$ --
these values give the same mean velocity, but the   string density is closer to
the Bennett-Bouchet value, though is consistent within the quoted error bars).

In an open, $\Omega  < 1$ FRW space-time, for normal types  of matter (ie dust
or radiation), $H$ does not decay like $t^{-1}$. Ignoring the curvature terms,
the solution of (\ref{vevolutioneqn}) for which $v=$constant is inconsistent
with $L \propto t$ from (\ref{vevolutioneqn}), and inconsistent with $L \propto
H^{-1}$ from (\ref{Levolutioneqn}). As noted in Ref. \cite{Martins}, it does
not appear possible to find a scaling solution in an open universe.

We have numerically solved the evolution equations
(\ref{Levolutioneqn}-\ref{vevolutioneqn}) with the expansion scale factor given
by (\ref{dustfactor}). We choose the initial conditions for $L/t$ and $v$ to be
given by the $\Omega =1$ dust era scaling solution of $v = 0.61$ and $L/t =
0.53$. Our reasoning is that at early times, when the scale factor behaves as
$a\propto t^{2/3}$,   the evolution is indistinguishable from an $\Omega =1$
space-time.  Only at late times is the effect of the curvature-dominated
expansion important.  Similarily, we assume that the coefficients $\kappa$ and
$\tilde c$, describing the small scale structure and chopping efficiency, are
unchanged from their dust-era values. This is somewhat unrealistic, since these
parameters differ even between the radiation- and dust-eras. Up to
radiation-matter equality, however, work by MS has shown  that the effect on
the evolution is dominated by the change in the expansion rate rather than the
shift in the parameters. Hence,  we expect our model to be reliable for the
observationally allowed values of $\Omega$  as we do not follow the evolution
too far beyond matter-curvature equality.

Sample results are displayed in Fig.  \ref{figure1}. We see that the rms
velocity and string density decrease rapidly at late times when the curvature
begins to dominate the expansion. For $\Omega  = 0.2$, the velocity drops to $v
= 0.41$, and the length scale grows to $L/t = 0.63$. By letting $\Omega \to 1$
in equations (\ref{Levolutioneqn}-\ref{vevolutioneqn}), we find final values $v
= 0.44$ and $L/t = 0.64$. Hence, the rapid, curvature-dominated expansion is the
main cause of the departure from scaling, since  the spatial curvature terms
contribute only a $\lesssim 10\%$ effect to the evolution. Consequently, our
results depend only very weakly on the Hubble parameter, $h$.  As pointed out by
Martins, the energy density in long strings is actually growing relative to the
background cosmological fluid. At sufficiently late times, the strings will come
to dominate the energy density of the universe. This deviation from the scaling
solutions should have an important effect on the large angle CMB anisotropy due
to cosmic strings.

\section{CMB Anisotropy}
\label{seccmb}

While it is beyond our means to simulate the evolution of cosmic strings in an
$\Omega  < 1$, dust-dominated FRW space-time at present, we may nevertheless
adapt our model for the quantitative evolution of a string network to estimate
the amplitude of CMB temperature anisotropy induced by cosmic strings.

We would like to determine the COBE-smoothed  rms temperature anisotropy due to
cosmic strings in an $\Omega  < 1$ cosmology.  Hence, we must compute 
$C(0^\circ,10^\circ)$, the $0^\circ$ angular separation correlation
function smoothed over $10^\circ$ in the manner of COBE.  To do so, we will
make the following simplifying assumptions:
\begin{enumerate}
\item
The large angle CMB anisotropy is due to the gravitational perturbations
caused by cosmic strings along the line of sight out to the surface of last
scattering.
\item
The mean, observer-averaged angular correlation function may be written as the
sum of the contributions by strings located in the time interval $[t,t+\delta
t]$; the contribution due to strings separated by an interval larger than the
characteristic time scale, $\delta t \gtrsim L$, is negligible.
\item
The effect of the negative spatial curvature in the open universe is to shift 
temperature anisotropy correlations to smaller angular scales than in a
spatially flat universe.
\item
The mean rms temperature anisotropy contributed in a time interval $\delta t$ is
proportional to the density of strings present, and the mean string velocity
during that interval. This is similar to Perivolaropoulos' model
\cite{PeriModel} in which the CMB anisotropy is a superposition of random
impulses due to the Kaiser-Stebbins effect \cite{KaiserStebbins},  for which
$\delta T \propto 8\pi G\mu v$, for each long string present.
\end{enumerate}
Given the first two assumptions, the correlation function may be
written as
\begin{equation}
C(\theta, z_{ls}) = \int_0^{z_{ls}} dz \, C(\theta)_{,z} .
\end{equation}
Here, $C(\theta,z_{ls})$ is the temperature correlation function contributed by
strings out to the redshift  of last scattering, $z_{ls}$. The function
$C(\theta,z)$ has been tabulated from the numerical simulation of CMB
anisotropy induced by cosmic strings in an $\Omega =1$, dust-dominated FRW
space-time (see Fig. 3 of \cite{AllenEtAl} where we observe that the dominant
contribution to the rms anisotropy for $\Omega  = 1$ occurs within a redshift
$z \lesssim 10$).  The function $C(\theta)_{,z}$ is obtained empirically by
differentiating $C(\theta,z)$. This procedure does not rely on assumption 2
above. However, in order to interpret $C(\theta)_{,z}$ as the contribution to
the angular power spectrum due to strings in the interval $[z, z+\delta z]$, we
must restrict our use to a time resolution $\delta t \gtrsim L$, greater than
the characteristic time scale.

The negatively curved spatial sections of the open FRW space-time lead to a
generic suppression of large-angle correlations. (See Ref.
\cite{StebbinsCaldwell} and references therein for more discussion.) We may
understand this effect by considering  that an object with angular size
$\theta_{-1}$ at a redshift $z$ from an observer in an open FRW
space-time subtends a smaller angle $\theta_{-1} < \theta_0$  than the angle of
the same object from the same redshift in a spatially flat FRW space-time. (The
subscripts $0,\,-1$ refer to the sign of the spatial curvature.) We may express
this relationship between the angles subtended as
\begin{equation}
\theta_{-1} \equiv f(\theta_{0},z,\Omega)   
= 2 \sin^{-1}\Big[\sin{\theta_0 \over 2}{\Omega^2 (1 + z - \sqrt{1+z})
\over \Omega z + (2 - \Omega)(1 - \sqrt{1 + \Omega z})} \Big]
\label{angleq}
\end{equation}
As a result, $\theta_{-1} \le \theta_{0}$ for all $z \ge 0$ and $\Omega \le 1$.
In the limit $\Omega \to 1$ or $z \to 0$, equation (\ref{angleq}) reduces to the
identity, with $\theta_0 = \theta_{-1}$. In order to include the effect of the
geometry on the temperature anisotropy correlation function in an open universe,
we write
\begin{equation}
{d C_\ell \over dz}(\Omega) =  2 \pi \int_0^\pi d(\cos\theta_{-1}) \,
P_\ell(\cos\theta_{-1}) \, C(\theta_0)_{,z}.
\label{geoeffect}
\end{equation}
By shifting the argument of the Legendre polynomial to smaller angles,
correlations on a particular angular scale are associated with a larger
$\ell-$mode in an open than in a flat FRW space-time.
  
To implement our final assumption above, we model the effect of the  
curvature-dominated expansion on the correlation function by weighting the
contribution at different redshifts using our model of quantitative string
evolution:
\begin{equation}
C_\ell(\Omega) = \int_0^{z_{ls}}  dz \, \Big[
{\rho_L(\Omega,z) \over \rho_L(\Omega=1,z)}
{v^2(\Omega,z) \over v^2(\Omega=1,z)}  \Big]  
\, {d C_\ell \over dz}(\Omega).
\label{cmbaniso}
\end{equation}
Hence, the moments of the correlation function, which is proportional to
$(\delta T)^2$, are weighted by two powers of the string velocity relative to
the $\Omega = 1$ value. We model the contribution of the $N$ long strings in
each volume to the temperature amplitude as $\sqrt{N}$, so that only one factor
of the string density relative to the $\Omega =1$ value is included above.  The
functional dependence of $\rho_L$ and $v$ on the redshift for a given open
cosmology is obtained by integrating
(\ref{Levolutioneqn}-\ref{vevolutioneqn}).  Because these weights change on a
time scale comparable to or slower than $L$, assumption 2 is satisfied.

We foresee that the CMB anisotropy will be diminished due to the dilution of
the string density, the decrease in mean velocity, and the negative spatial
curvature in an $\Omega < 1$ universe. The geometric effect due to the negative
spatial curvature, in equation (\ref{geoeffect}), will lead to a decrease in
the amplitude of the anisotropies generated at distances to the observer which
are large compared to the curvature length scale. The dynamical effect due to
the late time evolution of the string network, in equation (\ref{cmbaniso}),
will lead to a decrease in the amplitude of the anisotropies generated at late
times. The result is an overall decrease in the amplitude of the CMB anisotropy
spectrum for a given mass-per-unit-length.

We may estimate the normalization of  the cosmic string mass per unit
length $\mu$ in an open universe by comparing the observations of COBE-DMR
\cite{COBE} with our predictions. We carry out a procedure similar to that
given in \cite{AllenEtAl},  computing the smoothed autocorrelation function 
\begin{equation}
C(0^\circ,10^\circ,\Omega) \equiv 
\sum_{\ell=2}^\infty {2 \ell + 1 \over 4 \pi} \, 
|G_\ell|^2 |W_\ell(7^\circ)|^2 \, C_\ell(\Omega).
\end{equation}
Here, we smooth the temperature pattern first with the average DMR beam model
window function $G_\ell$ (tabulated values are given in \cite{WrightG}) which is
approximately a $7^\circ$ beam, and second with a $7^\circ$ FWHM Gaussian window
function $W_\ell(7^\circ)$ for an effective smoothing of $10^\circ$. Thus, we
find for the case $\Omega = 0.2$,  $G \mu = 1.7^{+0.6}_{-0.3}\times 10^{-6}$. 
The effect of the spatial geometry on the smoothed autocorrelation function is
only $\sim 20 \%$ for $\Omega = 0.2$; the dilution of the string  density and
the decrease in the mean velocity due to the rapid expansion are the main causes
of the change in the rms anisotropy amplitude. We have rescaled the error bars
assessed in \cite{AllenEtAl}, assigning no errors due to the crudeness of our
model. For $\Omega \sim 1$ this seems reasonable; for low-$\Omega$ we
underestimate our uncertainty in the normalization. The empirical formula for
the CMB-normalization of the string mass per unit length
\begin{equation}
G \mu(\Omega) = G \mu(\Omega=1) \Omega^{-0.3} =
1.05{\,}^{+0.35}_{-0.20} \times 10^{-6} \, \Omega^{-0.3}
\label{munorm}
\end{equation}
fits our results to within $5\%$ for $0.1 \le \Omega \le 1$.  We stress that
our estimate of the normalization is valid, within the above mentioned error
bars, insofar as Allen {\it et al} \cite{AllenEtAl} have accurately simulated
the large angle CMB anisotropy induced by realistic cosmic strings. This is the
the third main result of this paper. 

We take this opportunity to comment on the effect of an open universe on the
small angular scale CMB anisotropy induced by cosmic strings. Although no firm
predictions of the high$-\ell$ $C_\ell$ spectrum have been made to present,
recent work \cite{Allen97,PST97} has shed light on the qualitative features of
the spectrum. Based on numerical simulations, they observe a feature near
$\ell  \sim 100$ attributed to the decay of vector perturbations smaller than
the horizon scale on the surface of last scattering. For higher $\ell$, there
is a single, low, broad feature  (as opposed to the secondary oscillations
predicted in inflationary scenarios) in the range $\ell \sim 400 - 600$, as
conjectured by Magueijo {\it et al} \cite{SmallCMB}. In an open universe, the
apparent size of fluctuations near the surface of last scattering shift to
smaller angles as $\theta \propto \Omega^{1/2}$.  Hence, we expect the location
of the feature due to the decay of the vector perturbations to shift as $\ell
\sim 100 \, \Omega^{-1/2}$ towards smaller angular scales. 

We end these comments on the small angular scale spectrum by adding that MS
have shown that the transient in the evolution of the long string density and
velocity across the radiation-matter transition, observed in the 
Bennett-Bouchet numerical simulations \cite{BennettBouchet,BennettBouchetProc},
may last as late as $\sim 10^3 t_{eq}$ (see Fig. 18c,d of
\cite{MartinsShellard}).  In particular, the ratio $\rho_L/\rho_{crit}$, which
is higher in the radiation era, does not settle down to the matter  era scaling
value until $\sim 10^3 t_{eq}$, and the evolution of the mean velocity displays
a peak near $\sim 30 t_{eq}$ before reaching the matter era value. For low
values  of $\Omega$ and $h$, the redshift of radiation-matter equality
approaches last scattering, so that this transient may have an important effect
on the small angle CMB anisotropy generated near the surface of last scattering
\cite{ShellardCommunique}.

\section{Large Scale Structure}
\label{seclss}

Finally, we consider the large scale structure formation scenario with cosmic
strings. We will examine both the HDM and CDM scenarios, by adapting the
methods of Ref. \cite{AlbrechtStebbins} to estimate the power spectrum of
density fluctuations produced by cosmic strings. While the effect of
low-$\Omega$ on these string scenarios has been examined previously by Mahonen
{\it et al} \cite{MahonenMHM} and Ferreira \cite{Ferreira}, our contribution
will be the effect of the quantitative string evolution and the normalization
of $\mu$ on the power spectrum. 

In the semi-analytic model of Albrecht and Stebbins, the power spectrum  of
density perturbations induced by cosmic strings in an $\Omega=1$ universe is
approximated by 
\begin{eqnarray}
&& P(k) = 16 \pi^2 (1 + z_{eq})^2 \mu^2 \int_{\tau_i}^\infty
|T(k;\tau')|^2 {\cal F}(k \xi/a) d\tau' \cr\cr
&& {\cal F}(k \xi/a) = {2 \over \pi^2}
{{ \beta^2 \Sigma}} {\chi^2 \over \xi^2}
\Big(1 + 2(k \chi/a)^2 \Big)^{-1} .
\label{powerspectrum}
\end{eqnarray}
In these equations, $a$ is the scale factor which evolves smoothly from
radiation- to dust-dominated expansion,  $\tau_i$ is the conformal time at
which the string network formed, and $T(k,\tau')$ is the transfer function for
the evolution of the causally-compensated perturbations (see equation 2 of
\cite{AlbrechtStebbins} and equations 5.23, 5.45 of
\cite{VeeraraghavanStebbins}), specific to either CDM or HDM. In the case of
HDM, $T(k,\tau')$ includes a term fit to numerical calculations of the damping
of perturbations by non-relativistic neutrinos.   

The parameters used in the Albrecht-Stebbins estimate of the cosmic string
power spectrum are given by
\begin{equation}
\xi \equiv (\rho_L / \mu)^{-1/2}, \qquad
\beta \equiv \langle v^2 \rangle^{1/2}, \qquad
\Sigma \equiv
{\mu_r \over \mu}\gamma_b \beta_b + {1 \over 2 \gamma_b \beta_b}
\Big({\mu_r^2 - \mu^2 \over \mu \mu_r}\Big)
\end{equation}
where $\chi$ is the curvature scale of wakes, $\beta_b$ is the macroscopic bulk
velocity of string, $\gamma_b = (1 - \beta_b^2)^{-1/2}$,  and $\mu_r$ is the
renormalized mass per unit length, which reflects the accumulation of small
scale structure on the string. The ``I-model'' developed by Albrecht and
Stebbins uses the following values of the parameters:
\begin{center} 
\begin{tabular}{c|ccccc}
era&$\xi/(a \tau)$&$\chi/\xi$&$\beta$&$\beta_b$&$\mu_r/\mu$ \\
\tableline
rad'n&$0.16$&$2.0$&$0.65$& $0.30$ & $1.9$ \\ \hline
dust&$0.16$&$2.0$&$0.61$& $0.15$ & $1.4$ \cr  
\end{tabular} 
\end{center}
We note that the values of $\beta$, $\beta_b$, and $\mu_r$ were taken from the
Allen-Shellard (AS) simulation  \cite{AllenShellard,AllenShellardProc}. The
values of $\xi$ and $\chi$, however, reflect an estimate based on the
Bennett-Bouchet (BB) \cite{BennettBouchet,BennettBouchetProc} and AS
simulations.  The radiation era values were used for the I-model to determine
the power spectrum in a spatially flat, $\Omega   = 1$ universe.

While the I-model closely resembles the AS and BB simulations, we might
hope to make an improvement by including the effect of the evolution of
the string network parameters across the transition from radiation- to
dust-dominated expansion \cite{ShellardCommunique}. As investigated by
MS, the ratio $L/t$ interpolates between the radiation- and
dust-dominated scaling values, whereas the mean velocity displays a
short burst during which the string network rapidly sheds loops.  (See
Fig. 18c,d of \cite{MartinsShellard}.) Hence, we make the
identifications $\xi = L$ and $\chi = 2 \xi$, and use the evolution
of $L/t$ to interpolate between radiation- and dust-era values of
$\mu_r$ and $\kappa$, and use $\beta$ to guide $\beta_b$. Note that the
radiation era values are $(\tilde c,\kappa) = (0.24,0.18)$ for the BB
simulation, and $(0.22,0.16)$ for the AS simulation.  Applying the
model of realistic network evolution to the power spectrum, we find
that for the same $\mu$, the only change is a $\sim 30\%$ boost in the
power for the AS values, which is consistent with the quoted
uncertainties on the parameters measured in the simulations.

To adapt the power spectrum for an open universe, we would like to use the
transfer function $T(k,\tau')$ appropriate for $\Omega < 1$. In the present
work, however, we will use the $\Omega=1$ transfer function, which should be
satisfactory on the scales of interest, $\lambda \lesssim 10^2$ Mpc. Because
perturbations do not grow as fast as $\delta \rho/\rho \propto a$  in a low
density universe we use the factor $g(\Omega)$ (defined in equation \ref{signorm0})
to modify the amplitude of present-day  perturbations \cite{GFactor}.
Hence, we obtain the power spectrum
(adapted from \cite{AlbrechtStebbins})
\begin{equation}
4 \pi k^3 P(k) = 
{4 \pi \Omega^2 h^4 \theta_1^2 \mu_6(\Omega)^2 k^4 {g(\Omega)^2} \over
1 + (\theta_2 k) + (\theta_3 k)^2 +
(\theta_4 k)^3 + (\theta_8 k)^4 + (\theta_6 k)^{\theta_7}}
\Big[{ 1 \over 1 + 1/(\theta_5 k)^2} \Big]^2
\label{stringpower}
\end{equation}
\begin{center} 
\begin{tabular*}{10cm}[]{c|c@{\extracolsep{\fill}}ccccccc}
model& $\theta_1$& $\theta_2$& $\theta_3$& $\theta_4$& 
$\theta_5$& $\theta_6$& $\theta_7$& $\theta_8$  \\ \hline 
HDM &6.8 & 4.7 & 4.4 & 1.55 & 2198 & 2.46 & 6.6 & 3.2  \\ 
CDM &6.8 & 4.7 & 4.4 & 1.55 & 2198 & 0 & 0 & 0  \\ 
\end{tabular*} 
\end{center}
where $k$ is measured in units $\Omega h^2/$Mpc and $\mu_6(\Omega)
\equiv G \mu(\Omega) \times 10^{6}$ obtained from (\ref{munorm}).
Numerical simulations of string seeded structure formation 
by Avelino \cite{Avelino}, based
on the Allen-Shellard simulation, find agreement with equation
(\ref{stringpower}) in a flat universe on the limited range of scales
accessible to the simulation.

Sample power spectra for various cosmological parameters, constructed
using (\ref{stringpower}), are shown in Fig. \ref{figure3}.  The string
mass per unit length $\mu$ in each of the curves has been determined by
the CMB normalization obtained from equation(\ref{munorm}) in section
\ref{seccmb}.  In the top panels, the power spectra for $h = 0.7$ and
$\Omega = 1.0,\,0.4,\,0.2$ are shown. For reference, the standard CDM
power spectrum \cite{MA} is also displayed. In the three descending
panels, the individual spectra are shown with the Peacock and Dodds
\cite{PeacockDodds} (PD) reconstruction of the linear power spectrum.
For $\Omega < 1$ the reconstructed spectrum has been scaled as $\propto
\Omega^{-0.3}$ (see equation 41 of \cite{PeacockDodds}) for
comparison.

We first consider structure formation by strings with HDM. Based on the
normalization of $\mu$ obtained by \cite{AllenEtAl}, we see from Fig.
\ref{figure3} that the power spectrum approximately fits the shape of
the PD spectrum on large scales.  As a gauge of the string+HDM model
for low-$\Omega$, we have computed the variance of the excess mass
fluctuation  in a ball of radius $R = 8 h^{-1}$ Mpc,
\begin{equation}
\sigma_8^2 = \int  |w(k R)|^2 4 \pi k^2 P(k) dk, \qquad
w(x)=3(\sin{x} - x\cos{x})/x^3 
\end{equation}
which is observed to be around unity
\cite{PeacockDodds,DavisPeebles,WhiteEtAl}.  
An excellent fit to our results is given by the empirical formula
\begin{eqnarray}
&& \sigma_8(\Omega, h) = 0.25(\pm{0.1}) \, \times 
\Big( \mu_6(\Omega)\,
{g(\Omega) \over \Omega }  \,
{\Gamma(1 + 2.6 \Gamma  - 1.6 \Gamma^2)
\over 1 +  (10 \Gamma)^{-2}} \Big)\cr\cr
&& {\rm with} \qquad g(\Omega) \equiv {5 \over 2} \Omega/
\Big[ 1 + {1 \over 2} \Omega + \Omega^{4/7}\Big]  
\label{signorm0}
\end{eqnarray}
which is valid to within $\sim 10\%$ for $0.1 \le \Omega \le 1$ and $0.4 < h <
0.8$. The error bars on $\sigma_8$ are estimated based on the quoted
uncertainties in the string parameters
\cite{BennettBouchet,BennettBouchetProc,AllenShellard,AllenShellardProc} and the
uncertainty in the CMB normalization of $\mu$ \cite{AllenEtAl} included in
$G\mu(\Omega)$, which we repeat are probably too small for low-$\Omega$. 
Evaluating (\ref{signorm0}) for various values of the cosmological parameters,
we predict $\sigma_8(1.0,0.5) = 0.25\pm{0.1}$  and $\sigma_8(0.2,0.5) =
0.05\pm{0.02}$. For $\Omega = 1$ the string+HDM scenario requires a modest boost
or bias in the power in order to achieve $\sigma_8 \sim 0.57 - 0.75$
\cite{PeacockDodds,WhiteEtAl}. These results are in agreement with past work by
Colombi \cite{ColombiThesis}, based on the Bennett-Bouchet simulations.  We
pause to note that the non-linear dynamics of wakes and filaments
\cite{AlbrechtStebbins,VVach,TVach,AguirreBrand,AvelinoShellard,MoessnerBrand,ZLBrand,SornBrand,VollickA,VollickB,MahonenHYMMA,MahonenHYMMB,Mahonen}
may produce such a bias sufficient to reproduce the observed clustering of
objects on large scales.  However, in an open universe the peak amplitude of
$k^3 P(k)$ drops and shifts to larger scales, so that some sort of
scale-dependent boost would be required to produce more power for $k \gtrsim 1\,
\Omega h^2/$Mpc. Hence, string+HDM in an open universe does not appear to be a
viable model for structure formation.

Structure formation by strings with CDM in a flat, $\Omega=1$ universe,  when
normalized on large scales, suffers from producing too much power on small
scales. As pointed out by \cite{Spergel,PenSpergel,Ferreira} this problem may
be overcome, as for standard CDM, in a low density, $\Omega < 1$ universe.
Examining Fig. \ref{figure3}, we see that the string+CDM power spectrum ``bends
over'' on small scales as we lower $\Omega$. Hence, for  $\Gamma \equiv \Omega
h \sim 0.1 - 0.2$, the spectrum approximately fits the shape  of the PD
reconstruction.  The variance of the mass fluctuation is given by the empirical
formula
\begin{equation}
\sigma_8(\Omega, h) = 0.9(\pm 0.5) \, \times 
\Big( \mu_6(\Omega) \,{g(\Omega) \over \Omega} \,
{\Gamma (1 - 0.36 \Gamma) 
\over 1 + (50 \Gamma)^{-2}}  \Big)
\label{signorm1}
\end{equation}
which is valid to within $\sim 10\%$ for $0.1 \le \Omega \le 1$ and $0.4 < h <
0.8$. Evaluating (\ref{signorm1}), we predict $\sigma_8(0.4,0.7) = 0.4\pm{0.2}$
and $\sigma_8(0.2,0.5) = 0.2\pm{0.1}$. Hence, for $\Gamma \sim 0.1 - 0.2$, the
range of values of the mass fluctuation excess falls well below the estimate of
$\sigma_8 = 0.6{}^{+32\%}_{-24\%}
\exp[({-0.36-0.31\Omega+0.28\Omega^2})\log\Omega]$
\cite{VianaLiddle} by a factor of $\sim 2 - 4$. Within the uncertainties quoted
in (\ref{signorm1}), a bias as low as $b \sim 1.5$ may be tolerated. Recent work
by Sornborger {\it et al} \cite{SornBrand} on the structure of cosmic string
wakes has shown that the ratio of the baryon to CDM density in wakes is
enhanced. For a single wake formed near radiation-matter equality, the baryon
enhancement at late times is $\sim 2.4$ in a region of thickness $\sim 0.3$Mpc. 
These results, which suggest that structure formation by strings is biased,
complement our conclusion that the string+CDM model may be a viable candidate
for the formation of large scale structure in an open universe.

\section{Cosmological Constant} \label{seccosmoconst}

In this section we briefly consider the effect of a cosmological 
constant on the cosmic string scenario. The background cosmology in this
case is a spatially flat, FRW space-time with a cosmological fluid
composed of vacuum- and matter-components such that
$\Omega_{m} + \Omega_{\Lambda} = 1$. The expansion
scale factor is given by the expression
\begin{equation}
a(t) = \Big[ {1 - \Omega_\Lambda \over \Omega_\Lambda} 
\sinh^2\Big( {3 \over 2} H_o t \sqrt{\Omega_\Lambda}\Big) \Big]^{1/3}
\end{equation}
where $H_o,\,\Omega_\Lambda$ are the present-day Hubble constant and
vacuum-matter density parameter.  We may now follow a similar procedure as
outlined in section \ref{secevol} to study the evolution of the long string
length scale $L$ and velocity $v$ by taking the spatially flat, $\Omega \to 1$
limit in equations (\ref{Levolutioneqn}-\ref{vevolutioneqn}).  In this case we
find that for a comparable matter density as in an open universe, the dilution
of the string energy density and the damping of string motion is much weaker in
the cosmological constant universe.  Note that the argument of the $\sinh$ in
the scale factor, evaluated at the present-day, is ${1\over 2}\log|(1 +
\sqrt{\Omega_\Lambda})/(1 - \sqrt{\Omega_\Lambda})|$. Hence, for
small-$\Omega_\Lambda$ the scale factor behaves to leading order as $a(t) \sim
t^{2/3}$, just as for matter-dominated expansion.  Only when $\Omega_\Lambda \to
1$ are the effects of the exponential expansion important, damping the string
motion. For example, in the case of $\Omega_m = 0.3$, the ratio $L/t$ is only
$\sim 5\%$ larger and the velocity is only $\sim 5\%$ smaller than the $\Omega_m
= 1$, spatially flat value. For the open universe with $\Omega = 0.3$, the ratio
$L/t$ has grown by $\sim 15\%$ and the velocity has dropped by $\sim 30\%$ from
their $\Omega=1$ values.

We may estimate the CMB normalization of the mass per unit length as a 
function of $\Omega_m$ following the methods of section \ref{seccmb}. However,
there is no correction for the geometry, since the spatial sections are
flat. Hence, we find the empirical formula
\begin{equation}
G \mu(\Omega_m)_\Lambda = 1.05{}^{+0.35}_{-0.20}
\times 10^{-6} \, \Omega_m^{-0.05}
\label{munorm2}
\end{equation}
fits our results to within $5\%$ for $0.1 \le \Omega_m \le 1$.  The subscript
$\Lambda$ is used to differentiate the above normalization from the case of an
open universe, in equation (\ref{munorm}). We see that the normalization is
relatively insensitive to the presence of a cosmological constant.

Finally, we may consider the properties of the cosmic string scenario for
structure formation with CDM in the presence of a cosmological constant. We may
adapt equation (\ref{stringpower}) for the string+CDM power spectrum by setting
$\Omega  = \Omega_m$ and using the appropriate growth factor \cite{GFactor}. 
The variance of the mass  excess on length scales $R = 8 h^{-1}$Mpc is fit by
the empirical formula
\begin{eqnarray}
&& \sigma_8(\Omega_m,h) = 0.9 (\pm 0.5) \,  \times 
\Big( \mu_6(\Omega) \, {g(\Omega_m,\Omega_\Lambda) \over \Omega_m}
\, {\Gamma (1 - 0.36 \Gamma)
\over 1 + (50 \Gamma)^{-2}} \Big)
\cr\cr
&& {\rm with} \qquad
g(\Omega_m,\Omega_\Lambda) \equiv {5 \over 2} \Omega_m /
\Big[\Omega_m^{4/7} - \Omega_\Lambda + 
(1  + {1 \over 2} \Omega_m)(1 + {1 \over 70} \Omega_\Lambda)\Big]
\label{signorm2}
\end{eqnarray}
which is valid to within $\sim 10\%$ for $0.1 \le \Omega \le 1$ and $0.4 < h <
0.8$.  We find that the amplitude of the string+CDM power spectrum with a
cosmological constant is higher than in an open universe with the same matter
density, as demonstrated in Fig. \ref{figure3}. Evaluating (\ref{signorm2}), we
predict $\sigma_8(0.2,0.5) = 0.3\pm{0.2}$ and $\sigma_8(0.4,0.7) = 0.5\pm{0.3}$.
Comparing to observations, based on the estimate 
$\sigma_8 = 0.6{}^{+32\%}_{-24\%}
\exp[({-0.59-0.16\Omega+0.06\Omega^2})\log\Omega]$
\cite{VianaLiddle} for a spatially flat universe, we find that
a  slightly lower bias than in an open universe, $b \sim 1.5 - 4$, is required.
Hence, the string+$\Lambda$CDM scenario may be viable if the strings generate a
sufficient boost to explain the biased clustering on $8 h^{-1}$Mpc scales.

\section{Conclusion}
\label{secend}

In this paper we have laid out many of the tools necessary to study cosmic
strings in an open universe.  We have first derived the equations of motion and
energy conservation in an $\Omega < 1$ FRW space-time.  We have extended the MS
model of quantitative string evolution \cite{MartinsShellard} to the case of an
open, $\Omega  < 1$ universe. We believe this extrapolation is reasonable for
the range of values of $\Omega$ of interest.  We have found that with the onset
of curvature dominated expansion, the long string energy density and mean
velocity decay rapidly.  We have shown that the resulting effect on the large
angle CMB temperature fluctuations induced by cosmic strings is a lower level
of anisotropy than in a critical, $\Omega = 1$ universe, for the same $\mu$. 
Constructing a semi-analytic model for the generation of CMB anisotropy in an
open universe, based in part on the AS numerical simulation
\cite{AllenShellard,AllenShellardProc}, we found that comparison with the
COBE-DMR observations \cite{COBE} leads to a higher normalization of the cosmic
string mass per unit length. To the extent that the CMB anisotropy induced by
realistic cosmic strings has been accurately simulated in Ref.
\cite{AllenEtAl}, we believe our results, equations
(\ref{munorm},\ref{munorm2}), are reliable within the errors discussed.  The
new normalization of $\mu$, the first estimate of the normalization of $\mu$ in
a low density universe (as far as we are aware), is consistent with all other
observational constraints on cosmic strings, including the bound on a
stochastic gravitational wave background arising from pulsar timing
\cite{CaldwellBattyeShellard}.

Finally, we have demonstrated the effect of an open, $\Omega  < 1$ universe on
the power spectrum of density fluctuations produced by cosmic strings with HDM
and CDM.  As we mentioned in section \ref{secintro}, the power spectrum $P(k)$
does not completely specify the cosmic string structure formation scenario.
Fluctuations generated by string wakes and filaments are non-gaussian, so that
knowledge of $P(k)$ alone is insufficient to specify all the properties of the
density field.  Although the linear power spectrum (\ref{stringpower}) is in
agreement with the results of Avelino \cite{Avelino} and Colombi
\cite{ColombiThesis} on a limited range of scales, we are unable to make finely
detailed comparisons with observations without more knowledge of the
distribution of cosmic string seeded density perturbations. For example, it is
not clear whether the estimates of the rms linear fluctuation in the mass
distribution \cite{PeacockDodds,WhiteEtAl} obtained from the various galaxy
redshift surveys, which depend strongly on the gaussianity of the initial
density field, are directly applicable to a theory with a non-gaussian
fluctuation spectrum. Nevertheless, we have found that the string+CDM spectrum
fits the shape of the PD reconstruction of the linear power spectrum
\cite{PeacockDodds} for cosmological parameters in the range $\Gamma \sim 0.1 -
0.2$. We have computed the variance of the mass fluctuation in a sphere of
radius $R = 8 \, h^{-1}$Mpc, requiring a bias $b \gtrsim 2$ for consistency with
the inferred $\sigma_8$ of the linear density field.  In the case of a
cosmological constant, a slightly lower bias is required than for an open
universe string+CDM spectrum with the same matter density. These findings are
similar to Ref. \cite{Ferreira}, in which the product $b G\mu$ was estimated in
order to fit the string+CDM spectrum to the 1-in-6 IRAS QDOT survey
\cite{FeldmanEtAl}, and to Ref.  \cite{PenSpergel}, in which the effects of an
open universe on global defects, including global strings and textures, were
considered.  The results of Ref.  \cite{SornBrand} indicate that the density of
baryonic matter is enhanced in CDM wakes by a factor of $\sim 2.4$, suggesting
that a bias $b \sim 2$ may be possible.  It is clear that high resolution
simulations, as Ref. \cite{SornWake}, are necessary to further develop of the
cosmic string structure formation scenario.

The results presented in this paper provide excellent motivation to
continue investigation of the cosmic string scenario, which should be
possible with the equations of motion for strings and the normalization
of $\mu$ for $\Omega < 1$.  

\acknowledgements

We would like to thank Chung-Pei Ma, Paul Shellard, Andrew
Sornborger, and Albert Stebbins for
useful conversations. 
P.P.A. is funded by JNICT (Portugal) under
`Programa PRAXIS XXI' (grant no. PRAXIS XXI/BPD/9901/96).
The work of R.R.C. is supported by the DOE at
Penn (DOE-EY-76-C-02-3071). 
C.M. is funded by JNICT (Portugal) under
`Programa PRAXIS XXI' (grant no. PRAXIS XXI/BD/3321/94).

\eject

\begin{figure}
\caption{
The evolution of a circular cosmic string loop formed
at $t=t_{eq}$ with an initial radius $R_{loop}=10\,t_{eq}$, in a universe with
$\Omega=1,\, 0.2$ given by the solid and dotted lines respectively. 
The top panel shows the evolution of the radius in units of $t_{eq}$
versus $\log_{10}(a/a_{eq})$. The bottom two panels show the evolution
of the velocity. An expanded scale shows the first oscillations  
as the loop enters the horizon, after which we show only the maximum
velocity in each period of oscillation.
}
\label{figure0}
\end{figure}
 
\epsfxsize=2.5cm \epsfbox[0 700 100 800]{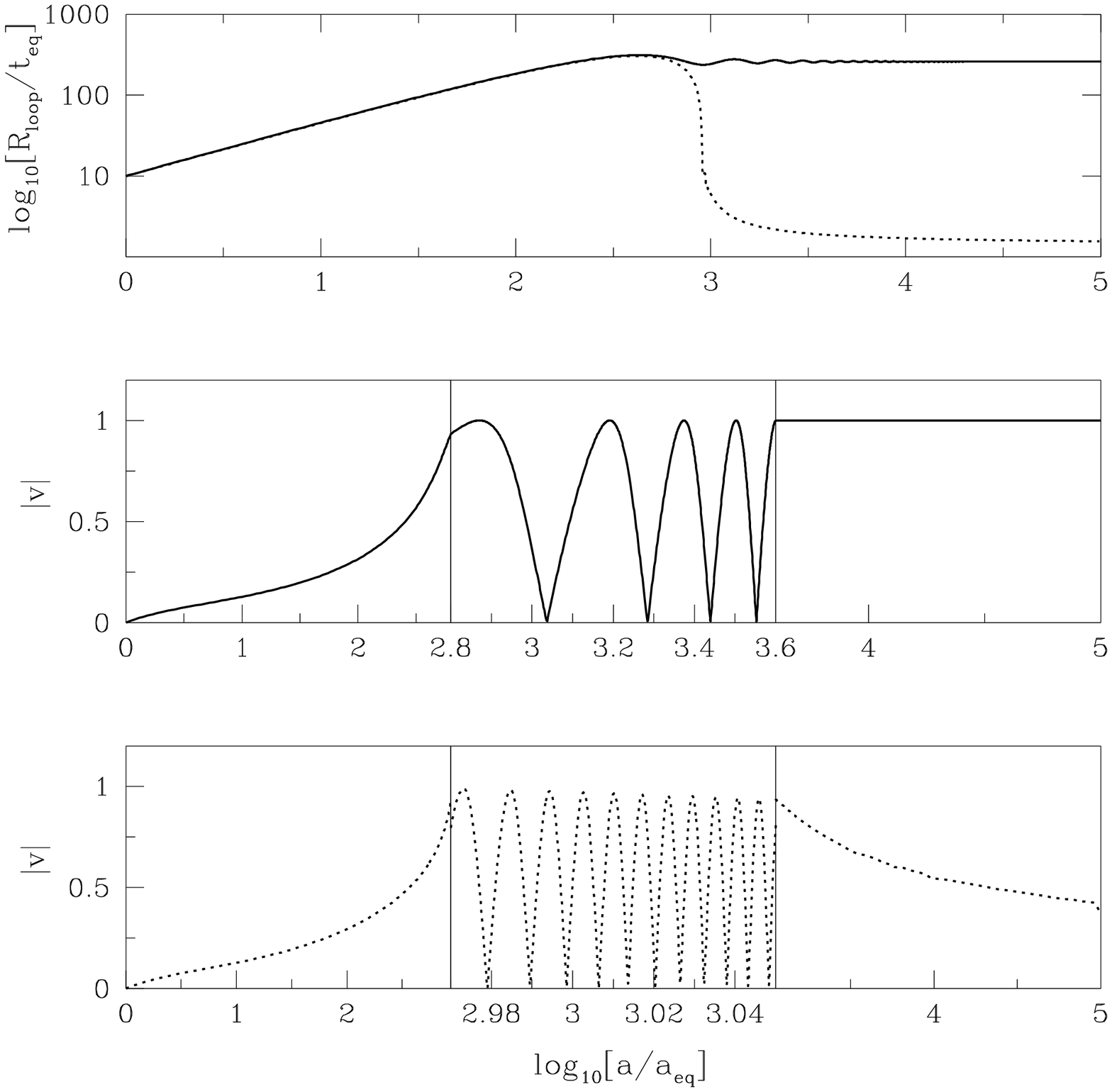}

\eject

\begin{figure}
\caption{
The evolution of the average string velocity and the characteristic
length scale of long strings in an open FRW space-time with $\Omega =
1.0,\, 0.6,\, 0.2$ given by the solid, dashed, and dotted curves. The
horizontal axis is the log of the cosmological time $t$. As the
expansion becomes curvature-dominated, the average velocity decays and
the characteristic string length scale grows. As a result, the number
of long strings in a box of linear dimension $t$ decreases, although
the string energy density relative to the background energy density
grows.
}
\label{figure1}
\end{figure}
 
\epsfxsize=2.5cm \epsfbox[0 700 100 800]{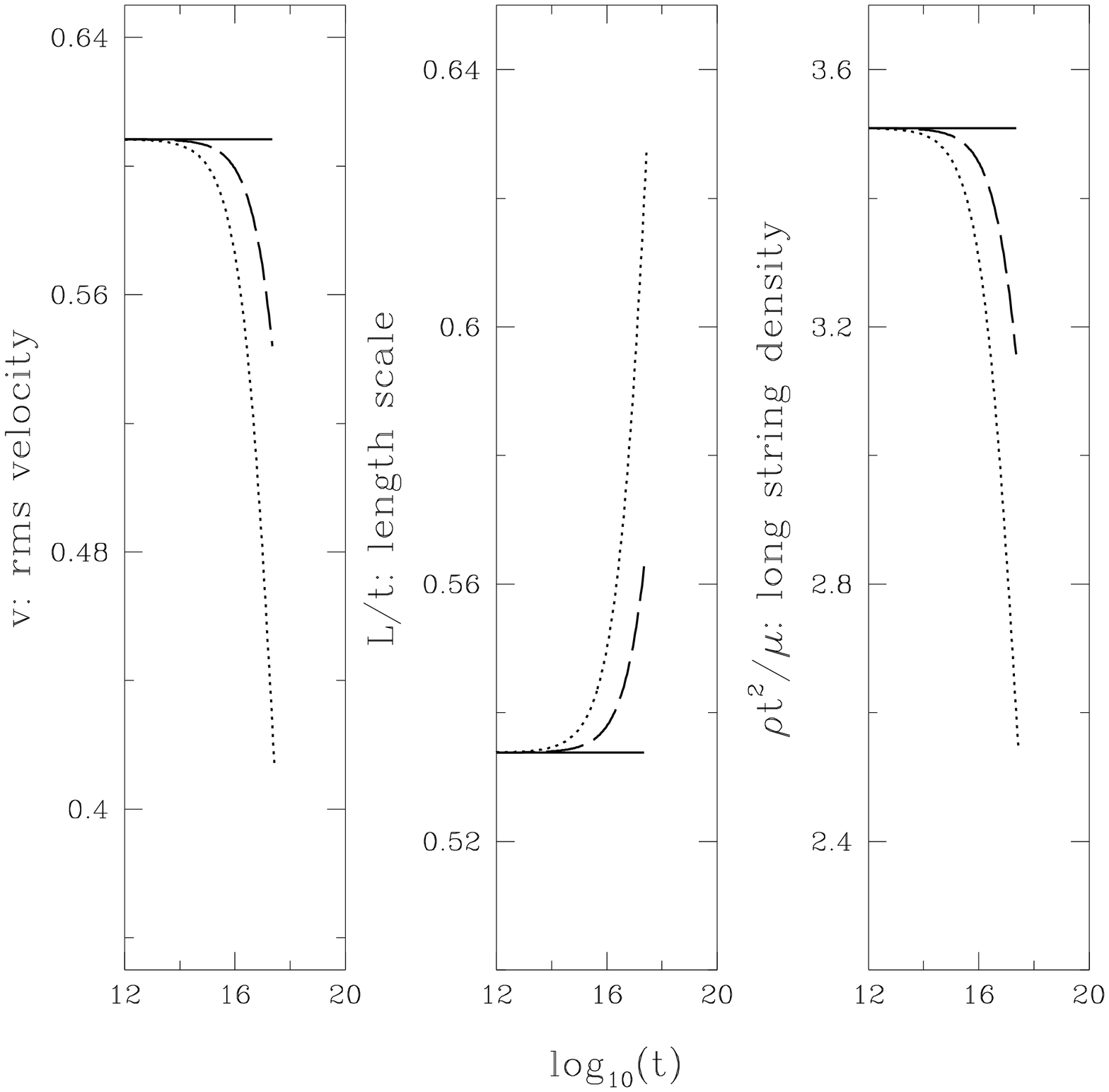}
 
\eject

\begin{figure}
\caption{
The CMB-normalized power spectrum $P(k)$ of density fluctuations
produced by cosmic strings with HDM (left) and CDM (right) are
presented for $\Omega = 1.0,\,0.4,\, 0.2$, given by the thick solid,
long-dashed, and short dashed curves.  For all cases, we have used
$h=0.7$. In the top panels, the thin solid line is the standard CDM
spectrum normalized to COBE following [34].  In the lower panels, the
data points are the PD reconstruction of the linear power spectrum,
with the amplitude rescaled $\propto \Omega^{-0.3}$.  In the bottom two
$\Omega < 1$ string+CDM panels, the thin solid line shows the
CMB-normalized power spectrum for the case of a cosmological constant
with the same matter density. A bias $b \sim 2-4$ is necessary to
obtain $\sigma_8 \sim 1$. In the presence of a cosmological
constant, a smaller bias is required.
}
\label{figure3}
\end{figure}

\epsfxsize=2.5cm \epsfbox[0 700 100 800]{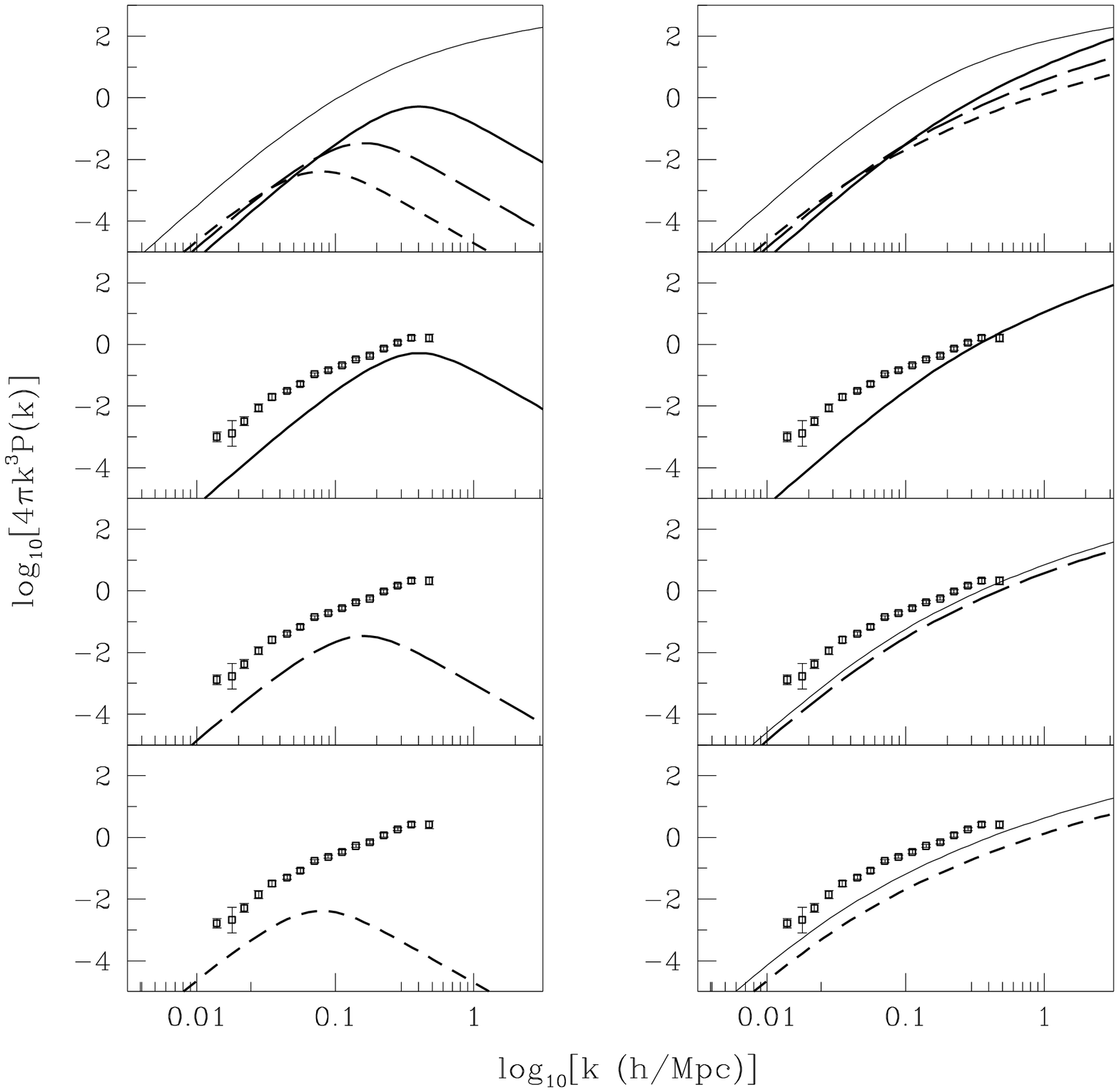}
  
\eject

\eject


\end{document}